\documentclass[12pt]{article}
\usepackage{graphicx}
\usepackage{hyperref}
\usepackage{amsmath}
\usepackage{url}
\usepackage{amssymb}
\usepackage{authblk}
\usepackage{amsfonts}
\usepackage{graphicx}
\usepackage{bm}
\usepackage{float}
\usepackage{todonotes}
\usepackage{caption}
\usepackage{subcaption}
\usepackage{textcomp}
\usepackage[natbibapa]{apacite}
\usepackage{setspace}
\usepackage[a4paper,top=1in,left=1in,right=1in,bottom=1in,includefoot,includehead]{geometry}

\usepackage[utf8]{inputenc}
\usepackage{microtype}

\newcommand{\indic}[1]{\mathbb{I}_{#1}}
\newcommand{\maxy}{\overline{y}}
\newcommand{\maxquant}{\overline{q}}
\newcommand{\vmaxy}{\overline{Y}}

\title{Pesticide concentration monitoring: investigating spatio-temporal patterns in left censored data.}

\author[1]{Clément Laroche}
\author[2]{Madalina Olteanu}
\author[2]{Fabrice Rossi}
\affil[1]{SAMM, EA 4543, Université Paris 1 Panthéon-Sorbonne, Paris, France}
\affil[2]{CEREMADE, UMR 7534, CNRS, Université Paris-Dauphine, PSL University, Paris, France}
\date{}

\begin{document}

\maketitle

\textbf{Abstract:} 
Monitoring pesticide concentration is very important for public authorities given the major concerns for environmental safety and the likelihood for increased public health risks. An important aspect of this process consists in locating abnormal signals, from a large amount of collected data. This kind of data is usually complex since it suffers from limits of quantification leading to left censored observations, and from the sampling procedure which is irregular in time and space across measuring stations. The present manuscript tackles precisely the issue of detecting spatio-temporal collective anomalies in pesticide concentration levels, and introduces a novel methodology for dealing with spatio-temporal heterogeneity. The latter combines a change-point detection procedure applied to the series of maximum daily values across all stations, and a clustering step aimed at a spatial segmentation of the stations. Limits of quantification are handled in the change-point procedure, by supposing an underlying left-censored parametric model, piece-wise stationary. Spatial segmentation takes into account the geographical conditions, and may be based on river network, wind directions, etc. Conditionally to the temporal segment and the spatial cluster, one may eventually analyse the data and identify contextual anomalies. The proposed procedure is illustrated in detail on a data set containing the prosulfocarb concentration levels in surface waters in Centre-Val de Loire region. 


\section{Introduction}

Monitoring the environmental pollution is of great interest for public authorities, important adverse health-effects being well documented nowadays \citep{khopkar2007,Marchant2018,NOUGADERE201432}. National health agencies are thus much concerned with monitoring ambient levels and quantifying the concentration of various pollutants in given environmental areas. 

At the same time, modelling environmental pollution data is a complex issue, due to several reasons, some intrinsic to the types of data under study, some specific to the data collection process implemented in different countries. Firstly, pollutant concentration levels are measured by sensors which have generally detection and quantification limits: the corresponding data are then left-censored. Secondly, the data is usually skewed to the right, with long tails hinting high concentrations. Thirdly, in numerous situations the data is irregularly sampled because of measurement practices, and is often multivariate, since various pollutant levels are monitored. Fourthly, pollution is monitored in various locations, each location possibly using different sensors, yielding a significant spatial heterogeneity. Notice that the last two problems are specific to some applications such as the one considered in the present paper. Indeed many countries have implemented very strict data collection protocols that ensure regular measurement rates on standardized sensors for a selection of pollutants\footnote{See for example the air quality data reported by UK-AIR \url{https://uk-air.defra.gov.uk/}.}. 

How to handle left-censored and right-skewed data is therefore one of the first aspects to consider when modelling environmental data. A rich literature has been developed on this topic during the last thirty years, and may be roughly divided into three categories of approaches: substitution methods (censored data is imputed using some values chosen \emph{a priori} or via a generative model), parametric methods (maximum likelihood estimates are computed under the hypothesis that the data comes from some log-normal, Weibull, Gamma, exponential, or other log-logistic distribution), and non-parametric methods (Kaplan-Meier or hazard function estimates). Detailed reviews of the various approaches are available  for instance in \citet{Authority2010,Hewett2007ACO,Mitra2008,Canales2018,Antweiler2008,Gillespie2010,shoari2018toward}.

The second aspect to consider is spatio-temporal heterogeneity. Air pollution data has received, for instance, a great deal of attention, and several modelling approaches have been proposed in the literature. Some are based on temporal regression models combined with kriging \citep{sampson2011,lindstrom2014flexible}, while others use latent variables and co-clustering approaches \citep{bouveyron2021co}. Nevertheless, these approaches do not include the fact that monitoring data is not normally distributed, and is usually left-censored. In the specific field of pesticide concentration monitoring, several recent papers address the spatio-temporal issue from an exploratory point of view \citep[see for instance][]{masia2016,figueiredo2021spatio,aznar2017spatio}. 

If one focuses specifically on temporal heterogeneity, a common approach to deal with it is to use a change-point based segmentation. Assuming the data is strictly stationary, conditionally to a (possibly unknown) number of change-points and their associated locations, change-point analysis aims at identifying the number of change-points (also known as\emph{breaks}), their locations, and the characteristics of the probability distribution within each temporal segment. Widely used in a variety of applications \citep{basseville1993detection, chen2012parametric, liu2017change, reeves2007review, levy2009detection}, and, in particular, for environmental pollution monitoring \citep{costa2016}, change-point detection is a reference technique for time series segmentation. The present manuscript relates to the offline framework, by supposing the full data has been recorded, and the segmentation is done posterior wise. For recent and detailed reviews of the offline change-point detection, the reader may refer to \citet{truong2020,bardet2020}. While the literature on change-point detection is abundant, applications to spatial data are somewhat limited. An early example of such method can be found in \citep{MAJUMDAR2005149} while recent advances in a setting close to ours are presented in \citep{doi:10.1080/07474946.2020.1826796}. As far as we know, none of the existing change-point detection method for spatial data applies to irregularly sampled and sparse data (on the temporal axis). 

The present manuscript tackles the issue of pesticide concentration monitoring, and introduces a new methodology which integrates both the specific left-censored distribution of the data, and the spatio-temporal context. The main goal is to identify contextual anomalies, both from a temporal and a spatial point of view. The proposed method builds on a parametric model for left-censored and right-skewed distributions, and combines it with a change-point detection step and a clustering step. 

Change-point detection is used for modelling temporal heterogeneity, by assuming a piece-wise stationary distribution on the series of maximum values, for a given time resolution. It produces temporal segments in which the pesticide concentrations are assumed to follow a stationary distribution. 

Clustering is then used for modelling the expected spatial homogeneity while integrating geographical constraints such as river networks, wind directions, etc. Indeed, as geological, terrain and climatic characteristics of an area can influence the dispersion of a chemical substance and on its potential use in the case of e.g. a pesticide, concentrations are expected to be somewhat correlated in small scale regions that are homogeneous in terms of influencing characteristics. Especially in the application presented here, which relates to the investigation of pollutants in surface waters, it is interesting to take into account the hydrographic structure of the region \citep[as in e.g.][]{doi:10.1080/07474946.2020.1826796}. Indeed, if a high concentration of a substance is detected at a certain point in time, traces of this substance should be found later downstream.
This hypothesis is accounted for by building clusters of measuring stations according to their proximities measured via the hydrographic network. 

Conditionally to the temporal segment detected by the change-point procedure, and to the spatial cluster detected by the clustering procedure, one may analyse the data and identify contextual anomalies. 

The rest of the manuscript is organised as follows: in Section \ref{section:data:model}, the generative model assumed for environmental pesticide monitoring data is described; the proposed method for estimating and handling this model from observed data is detailed in Section \ref{section:methods}; a detailed example on data collected by French authorities in Val de Loire region is fully illustrated in Sections \ref{section:data} and \ref{section:results}. 

\section{Data collection procedure and associated generative model}\label{section:data:model}
We study specifically in this paper a non homogeneous data collection process for pesticide use monitoring. It is represented by a generative model with two levels. The first level, a.k.a. the fine-grain level, consists of a network made of monitoring stations, where each station is associated to an irregularly sampled time-series. The second level, a.k.a. the coarse-grain level, summarises the maximum recorded values throughout the network, for a specified temporal resolution, and assumes a piece-wise stationary distribution. 

\subsection{Monitoring stations network}\label{subsection:graph}
We consider a network of monitoring stations used to collect concentration measurements at irregularly sampled instants. The stations are represented by an undirected graph $G=(V, E)$, which vertices $V=(v_i)_{1\leq i\leq N}$ are the monitoring stations and which weighted edges $E$ are links between stations that are directly comparable. The aim of the graph is to represent expert knowledge about expected measurement homogeneity. When two stations are connected in $G$, their measurements can be compared directly: a small edge weight assumes simultaneous measurements to be close, while a large one allows for significant differences. Shortest paths in the graph can be used to compare stations that are not directly connected, using the total weight of the paths to measure non homogeneity. This approach is inspired by methods developed for signal processing on graphs \citep{6494675}, but we use a dissimilarity based weighting rather than the classical similarity based one. 

This graph based representation is very flexible and can be used to model different types of spatial homogeneity. For instance, the focus of the present paper is the monitoring of water concentration of pesticides and thus dissimilarities between stations will be computed based on the network of rivers on which they are situated (see Section \ref{section:spaceclust}). Other modelling approaches may use a different graph considering for instance dominant wind directions relevant for air diffusion of pollutants. 

\subsection{Data collection}\label{subsection:data:collection}

Each station $v_i$ is supposed to be associated to a time series $(y_{ij},t_{ij})_{1\leq j\leq p_i}$, where $p_i$ is the number of sampled data points at $v_i$, and $y_{ij}$ is the concentration level of some pollutant at time $t_{ij}$. All measurements $y_{ij}$ are left-censored by some threshold $q_{ij}$, representing the quantification limit. Quantification limits depend on the machines used at each station and at each time instant, hence depend both on the station $v_i$ and on the collection instant $t_{ij}$. Furthermore, quantification limits are supposed to be known, fixed quantities. 

Summarising the above notations and hypotheses, a data set sampled from the stations network is given by a collection of measurements and associated quantification limits, and denoted
\begin{equation*}
 \mathcal{D}=\left(\left(y_{ij},t_{ij}, q_{ij}\right)_{1\leq j\leq p_i}\right)_{1\leq i\leq N}.   
\end{equation*}
Notice that in practical applications, we expect to have a rather small number of measurements for each station, i.e. to have small values for the $p_i$. In addition, we do not expect the measurement instants to be shared among the stations. See Section \ref{section:data-naiade} for examples.

From the complete representation of the data $\mathcal{D}$, one may derive an aggregated, coarser representation. First, an adapted temporal resolution for the phenomenon at study is selected. For instance, in the case of the present paper, a daily resolution is considered. Second, the selected resolution is used to build a time series of increasing instants $(\tau_k)_{1\leq k\leq K}$, at which at least one observation is available in the data collection. We denote $t_{ij}\in\tau_k$ the fact that the observation time $t_{ij}$ is compatible with $\tau_k$ at the specified resolution, e.g. that the observation $y_{ij}$ was made during the day $\tau_k$. 

Third, once $(\tau_k)_{1\leq k\leq K}$ has been computed,  one may introduce a coarse-grain, global series, summarising the maximum values recorded within the temporal resolution with
\begin{equation}
\maxy_k=\max\left\{y_{ij} \mid t_{ij}\in\tau_k\right\}.
\end{equation}
For instance, for a daily aggregation level, $\maxy_k$ is the largest value among all the measurements that took place during day $\tau_k$. Notice that $(\maxy_k)_{1\leq k\leq K}$ is left-censored as the consequence of the censoring of the underlying values. The quantification limit for $\maxy_k$ is denoted $\maxquant_k$, with
\begin{equation}
\maxquant_k=\max\left\{q_{ij}\mid t_{ij}\in\tau_k\right\}.   
\end{equation}
The coarse representation of $\mathcal{D}$ is then
\begin{equation}
\overline{\mathcal{D}}=\left(\maxy_k, \tau_k, \maxquant_k\right)_{1\leq k\leq K}.
\end{equation}

\subsection{A piece-wise stationary model for the coarse-grain time series}\label{subsection:pwsm}
In order to model the global use of the substance under monitoring, a piece-wise stationary generative model is introduced for the coarse data set $\overline{\mathcal{D}}$. The model is based on the following assumptions:
\begin{itemize}
    \item there are $L^*>0$ change-points producing $L^*+1$ stationary intervals defined by
\begin{equation*}
0=\eta_0^*<\eta_1^*<\ldots<\eta_{L^*}^*<\eta^*_{L^*+1}=K;    
\end{equation*}    
    \item the observations $(\maxy_k)_{1\leq k\leq K}$ are realisations of $K$ independent random variables $(\vmaxy_k)_{1\leq k\leq K}$;
    \item when $k\in [\eta^*_{l-1}+1, \eta^*_{l}]$, $\vmaxy_k$ is distributed according to a left-censored parametric distribution $Q$ with interval dependent parameters $\theta^*_l$ and a left-censoring threshold $\maxquant_k$, which is a known constant.
\end{itemize}
Notice that the model only accounts for the concentrations $\maxy_k$ but not for the instants and the quantification limits which are supposed deterministic quantities.

\section{Methods}\label{section:methods}
We are interested in finding anomalies in data collected according to this spatiotemporal model. The proposed methodology combines two different homogeneity models. The temporal aspect is based on the piece-wise stationary model proposed in Section \ref{subsection:pwsm}, while the spatial aspect is based on the graphical representation introduced in Section \ref{subsection:graph}. In a first step, we estimate the parameters of the temporal model. In a second step, the homogeneity assumptions represented by the graph of stations is used to detect stations with anomalous measurements with respect to close stations in a given stationary temporal segment. 

\subsection{Piece-wise stationary model estimation via change-point detection}\label{subsection:pelt}
The coarse-grained data $\overline{\mathcal{D}}$ is segmented using a change-point detection approach applied to the model introduced in Section \ref{subsection:pwsm}. 
Since both the number and the location of the change-points are unknown, we shall optimise a penalised cost function, and seek to estimate the number of change-points $L^*$, the change-point locations $\bm{\eta}^\star= (\eta^*_l)_{1\leq l\leq L^*}$, and the parameters $\bm{\theta}^\star= (\theta^*_l)_{1\leq l\leq L^*+1}$. 

\noindent The estimates write as
\begin{equation}
(\widehat{L}, \widehat{\bm{\eta}}, \widehat{\bm{\theta}}) = \arg \min_{L, \bm{\eta}, \bm{\theta}} \mathcal{C}(L,  \bm{\eta}, \bm{\theta};   \overline{\mathcal{D}}).
\end{equation}
The penalised cost is given by
\begin{equation}
\mathcal{C}(L,  \bm{\eta}, \bm{\theta};   \overline{\mathcal{D}})=\sum_{l=1}^{L+1}  -\ln \mathcal L_Q(\theta_l; \maxy_{\eta_{l-1}+1}, \ldots, \maxy_{\eta_l}) + \beta_K(L+1)D,
\end{equation}
where $\mathcal L_Q(\theta_l; \maxy_{\eta_{l-1}+1}, \ldots, \maxy_{\eta_l})$ is the likelihood of the $l$-th segment for the distribution $Q$, $\beta_K$ is the penalty to apply at the addition of new segment, and $D$ is the dimension of the parameter vectors $\theta_l$. Notice that the penalty $\beta_K$ depends on the size of the data $K$.

For fixed values of $L$ and of $\bm{\eta}$, $\mathcal{C}(L,  \bm{\eta}, \bm{\theta};   \overline{\mathcal{D}})$ is maximized by setting $\bm{\theta}$ to the maximum likelihood estimate (MLE), $\widehat{\bm{\theta}}_{MLE}(L, \bm{\theta})$.  Thus, the optimisation problem may be further written as 
\begin{equation}
(\widehat{L}, \widehat{\bm{\eta}}) = \arg \min_{L, \bm{\eta}} \mathcal{C}(L,  \bm{\eta}, \widehat{\bm{\theta}}_{MLE}(L, \bm{\eta});   \overline{\mathcal{D}}).
\end{equation}
The number of change-points $\widehat{L}$ and the associated locations $\widehat{\bm{\eta}}$ are obtained by applying the PELT procedure \citep{Killick2012}, which improves the optimal partitioning approach through a lower, linear complexity. The choice of the penalty term $\beta_K$ is driven by the CROPS algorithm \citep{haynes2017}, which computes all optimal segmentations as the penalty varies over some interval. Eventually, the final penalty value is selected using an elbow rule heuristic as proposed in \citep{lung2015homogeneity}: segmentation scores are plotted against their corresponding number of change-points $L$. One looks for the number of breaks $\hat L$ that minimizes the sums of squares of two linear models respectively fitted on the $L \geq \hat L$ and the $L \leq \hat L$. The penalty value associated to $\hat L$ is selected.   

\subsection{Spatial clustering}\label{section:spaceclust}
In any of stationary intervals identified in the previous step, the measurements are assumed to be consistent with the homogeneity assumptions represented by the graph $G=(V, E)$. A natural way of assessing the actual regularity of the measurements would be to use graph signal processing techniques \citep[see e.g.][]{8347162, 6494675}. However the irregular, unaligned, sparse and censored nature of the measurements at each station, prevents the use of such methods. The measurements are also incompatible with techniques designed to detect anomalous clusters in a graph \citep[see for instance][]{10.1214/10-AOS839}.

To circumvent this problem, we propose to leverage the graphical representation to build spatial aggregates and to assess homogeneity at this aggregated level. This corresponds to clustering the stations using the graph structure. We proceed as follows. 

Nodes of each connected component of the graph $G=(V, E)$ are clustered using a Ward hierarchical clustering method implemented on the shortest path distance computed from the edge weights. Those component specific hierarchies are combined in a global one in a greedy way. The initial global clustering of $V$ is obtained by assigning all vertices in a connected component to the same cluster. Subsequent levels of the global hierarchy are obtained by replacing the clusters of a connected component by the next refined level of the local hierarchy. At each step of the refinement, we select the component that reduce the most the inertia of the clustering. We use the standard definition of inertia given for the clustering $\mathcal{P}=(C_1,\ldots, C_M)$ by
\begin{equation}
    W(\mathcal{P}) = \sum_{m=1}^M \frac{1}{|C_m|}\sum_{v_i, v_j \in C_k}d^2_{ij},
\end{equation}
where $d^2_{ij}$ is the square of the shortest path distance in $G$ between vertices $v_i$ and $v_j$, and $|A|$ denotes the cardinality of set $A$. Clustering with a small inertia contain clusters that group close monitoring stations according to the graph $G$. 

To select the final clustering in the hierarchy, we use the same decision rule as \ref{subsection:pelt}. This time, the inertia of the clustering is plotted against the corresponding number of clusters $M$. We look for the number of breaks $M^*$ that minimizes the sums of squares of two linear models respectively fitted on the $M \geq M^*$ and the $M \leq M^*$.      

Notice that we rely on a simple graph clustering approach for two main reasons. Firstly, we do not expect graphs of monitoring stations to exhibit the specific characteristics of complex networks (such as very high degree vertices, small diameter, etc. see e.g. \citep{Newman2003GraphSurveySIAM}) that justify the use of techniques such as maximal modularity clustering \citep[see e.g.][for a survey]{FortunatoSurveyGraphs2010}. On the contrary, simpler approaches that interpret shortest paths weights as dissimilarities should be sufficient \citep[see e.g.][]{Schaeffer:COSREV2007}. Secondly, we work on relatively small graphs with even smaller connected components and we do not face computational issues associated to hierarchical clustering.   

\subsection{Anomaly detection}\label{section:anomaly}

Two types of anomalous clusters are targeted: either clusters with anomalous stations, or wholly anomalous clusters. Clusters containing anomalous stations are detected by studying the homogeneity of the measurements provided by the stations in a given spatial cluster. Anomalous clusters of stations are detected by simply pooling all measurements of each cluster to estimate the local use of the substance and detect large rates. We derive in this section two anomaly scores covering those cases.

For the first case, we need to assess the homogeneity of the measurements of the stations in a spatial cluster for a stationary time interval. As pointed out previously, the number of measurements provided by a single station is usually quite small, especially when we consider a single stationary interval. As a consequence classical distances between empirical distributions are not appropriate, mainly because the measurements of two stations do not have any value in common. Then the Kolmogorov-Smirnov statistics will be essentially driven by the number of observed values rather than the actual values, while other quantities, such as the Jensen-Shannon divergence, cannot be properly estimated (see appendix \ref{appendix:wasserstein}). For this reason, we propose to use the Wasserstein $w_1$ distance \citep{villani2009optimal}. For two discrete distributions on $\mathbb{R}$, it is expressed as the $L^1$-distance between their cumulative distribution functions and is therefore simple to compute. 

The measurement homogeneity of the clusters obtained in Section \ref{section:spaceclust} is therefore defined as the mean within cluster empirical Wasserstein average distance of a station measurements to the others. Denoting $C_m$ the $m$-th cluster and $|C_k|$ the number of stations present in $C_m$, $w_1(\mathbf{y}_i,\mathbf{y}_j)$ the empirical 1-Wasserstein distance between the data of stations $v_i$ and $v_j$, this quantity is expressed as   
\begin{equation}
    \bar{W}_k = \frac{1}{|C_m|(|C_m|-1)}\sum_{1 \leq j \leq |C_m|}\sum_{1 \leq i \leq |C_m|, i \neq j}w_1(\mathbf{y}_i,\mathbf{y}_j).
\end{equation}
The second type of potentially anomalous clusters are simply associated to the presence of quantified measurements and high values of concentration. Thus we estimate for each spatial cluster $C_m$ the parameters of distribution $Q$ (see Section \ref{subsection:pwsm}) on the pooled measurements obtained from all the stations of the cluster during the chosen stationary interval. From those parameters, we compute a statistics, denoted $\bar{I}_m$, used as a proxy for the intensity of the measurements (see Section \ref{subsection:anomalous} for an example).  Hence we consider a low concentration to be the normal case, but we do not define a threshold between normal clusters and abnormal ones. 

Each cluster $C_m$ is therefore characterised by two values $(\bar{W}_m, \bar{I}_m)$. To select potentially anomalous clusters, we use a multi-objective optimisation approach, considering that both characteristics are equally interesting. Following \citep{KIELING2002311}, we say that  $X_k = (\bar{W}_m, \bar{I}_m)$ is \emph{Pareto dominated by} $X_l = (\bar{W}_l,\bar{I}_l)$, and we write $X_m \prec X_l$ if and only if
\begin{equation*}
    \left((\bar{W}_m<\bar{W}_l)\text{ and }(\bar{I}_m\leq \bar{I}_l)\right)
    \text{ or }\left((\bar{W}_m \leq \bar{W}_l)\text{ and }(\bar{I}_m < \bar{I}_l)\right).
\end{equation*}
The level 1 Pareto optimal front is the set of maximal points for $\prec$. Level $b$ with $b>1$ is defined recursively as the optimal Pareto front computed for the set of points that do not belong to the optimal Pareto front of levels $1,\ldots,b-1$. Therefore clusters in the level 1 Pareto front are remarkable in the sense that there is no other cluster with higher heterogeneity and more extreme measurements. We define these clusters as anomalous. Pareto front and levels are evaluated using the Skyline algorithm \citep{914855,endres2015scalagon}.

\section{Data presentation}\label{section:data}

The methodology introduced in the above sections will be illustrated next using a case study on the prosulfocarb  concentration \citep{Prosulfocarb:NIH:url} in Centre-Val de Loire. This chemical compound is mainly used as a herbicide in field crops, with a typical period of active use in autumn. The monitoring of its concentrations in surface waters has been subject to increasing attention due to its aquatic ecotoxicology \citep{PPV,Prosulfocarb:PPDB}.

\subsection{Time period and geographical area selection}
\label{section:data-naiade}
Prosulfocarb usage was banned in France before 2007. A market re-authorisation was issued by the French Observatory on Pesticide Residues (now part of the ANSES\footnote{ANSES stands for \emph{Agence nationale de sécurité sanitaire de l’alimentation, de l’environnement et du travail}, i.e., French Agency for Food, Environmental and Occupational Health \& Safety.}) in 2009.
Since then, two modifications of the authorisation for use have been put in place, in November 2018 and in November 2019 respectively. Both changes consist in restrictions of use, one imposing specific equipment for application, the other restricting the application schedule in the presence of non-target crops next to the treated area. Motivated by these changes in regulation, the time period chosen for our study spans from January 1, 2007, to September 8, 2020.

Moreover, our study focuses on the geographical area of French Centre-Val de Loire region. Indeed, between 2009 and today, the annual mass of prosulfocarb sold in this region  exploded, making it rise from the 17th most sold substance in 2009 to the 4th in 2017 (see Figure \ref{fig:sale} in Appendix \ref{section:sale}). This region is also characterised by high concentrations of prosulfocarb target crops (such as the Beauce plains) (see Figure \ref{fig:crops} in Appendix \ref{section:crops}). 
Many target crops (cereal crops) are also concentrated in the region. These two elements combined guarantee a significant use of the product in this area.  
Thus, we expect significant variations in concentration values in this area during this period. 

Data about surface water quality in France is available from the French Biodiversity Agency \citep{Naiade}. We collected from the site the data selected above\footnote{Data exported in September 2020 using \url{http://www.naiades.eaufrance.fr/acces-donnees\#/physicochimie/resultats?debut=09-01-2007&fin=08-09-2020&regions=24&parametres=1092&fractions=23&supports=3&qualifications=1}}. 
These choices led to a data set $\mathcal D$ comprising 337 monitoring stations that performed 11,002 measurements. Each measurement is described by the monitoring station ID, the sampling date, the quantification limit (LOQ), and the concentration measurement value, if the concentration exceeds the LOQ. Indeed, as this is usually the case when measuring the concentration of a chemical substance \citep{Bernal2014,Currie1995}, the measurements are submitted to 
\begin{itemize}
    \item a limit of detection (LOD) which is the smallest concentration of the substance in a test sample that can be reliably distinguished from zero;
    \item a limit of quantification (LOQ) which is the smallest concentration of the substance in a test sample which can be measured reliably. 
\end{itemize}
The LOD is always lower than the LOQ and both quantities act as censoring values. In the data used in this work, the LOD is unknown: the left censoring phenomenon corresponds therefore to the LOQ of the measuring stations. When both limits are known, one can adapt the model proposed in Section \ref{subsection:data:collection} to take both of them into account: this would translate into a slightly more complex likelihood as the one derived in Appendix \ref{appendix:weibull} as we need to consider three cases (when the concentration is between 0 and the LOD, when the concentration is between the LOD and the LOQ, and finally when the concentration is observed and larger than the LOQ). 

\begin{figure}[htb]
  \centering
  \includegraphics[]{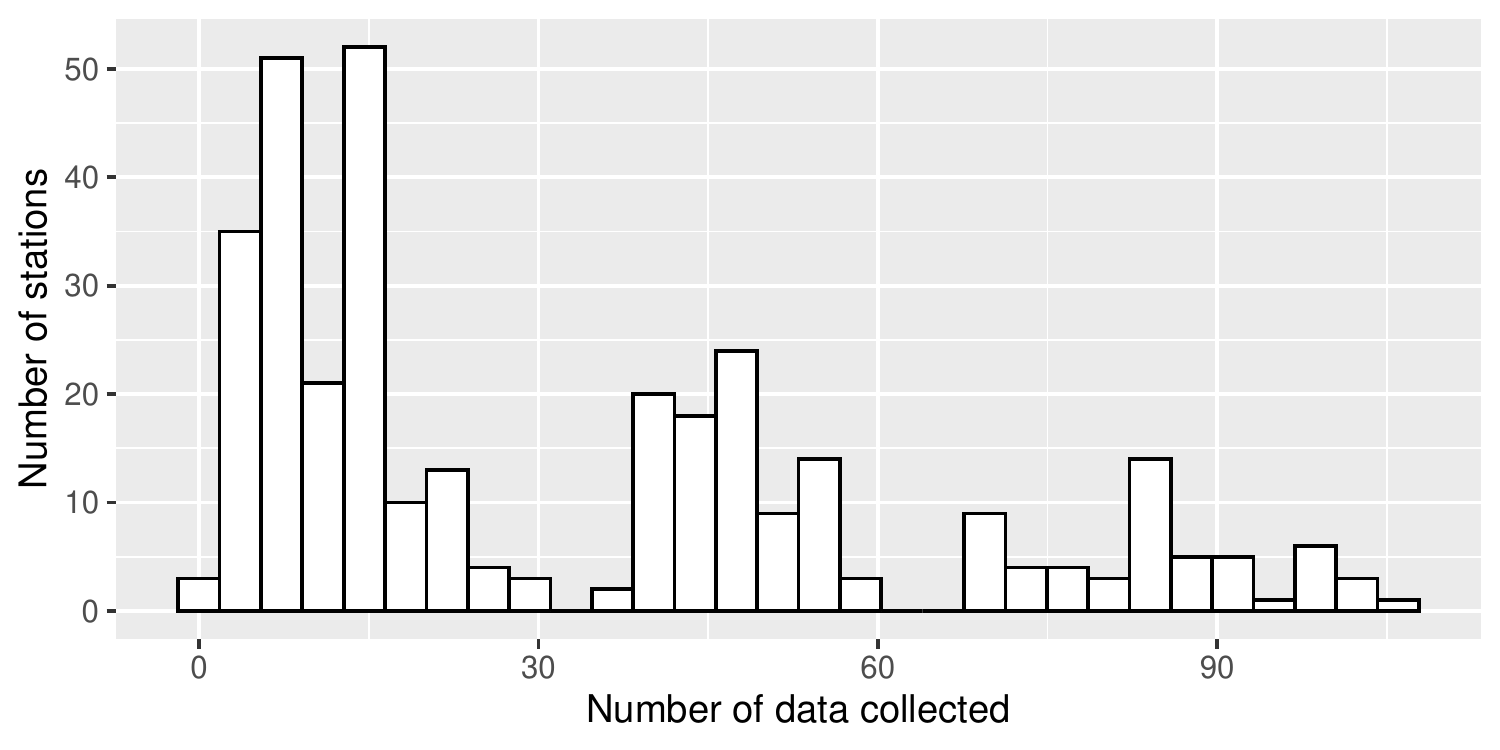}
  \caption{Distribution of the number of measurements per station.}
  \label{fig:histogram}
\end{figure}

Among the 11,002 recorded measurements during the period of interest, only 12.37\% were above the quantification limit. Figure \ref{fig:histogram} shows the distribution of the number of measurements per station: the mean and median number of samples collected by each monitoring station are respectively 32 and 19. This illustrates that sampling rates are different across stations, most of them making few measures, and the monitoring process is heterogeneous.

\begin{figure}[hbt]
  \centering
  \includegraphics[]{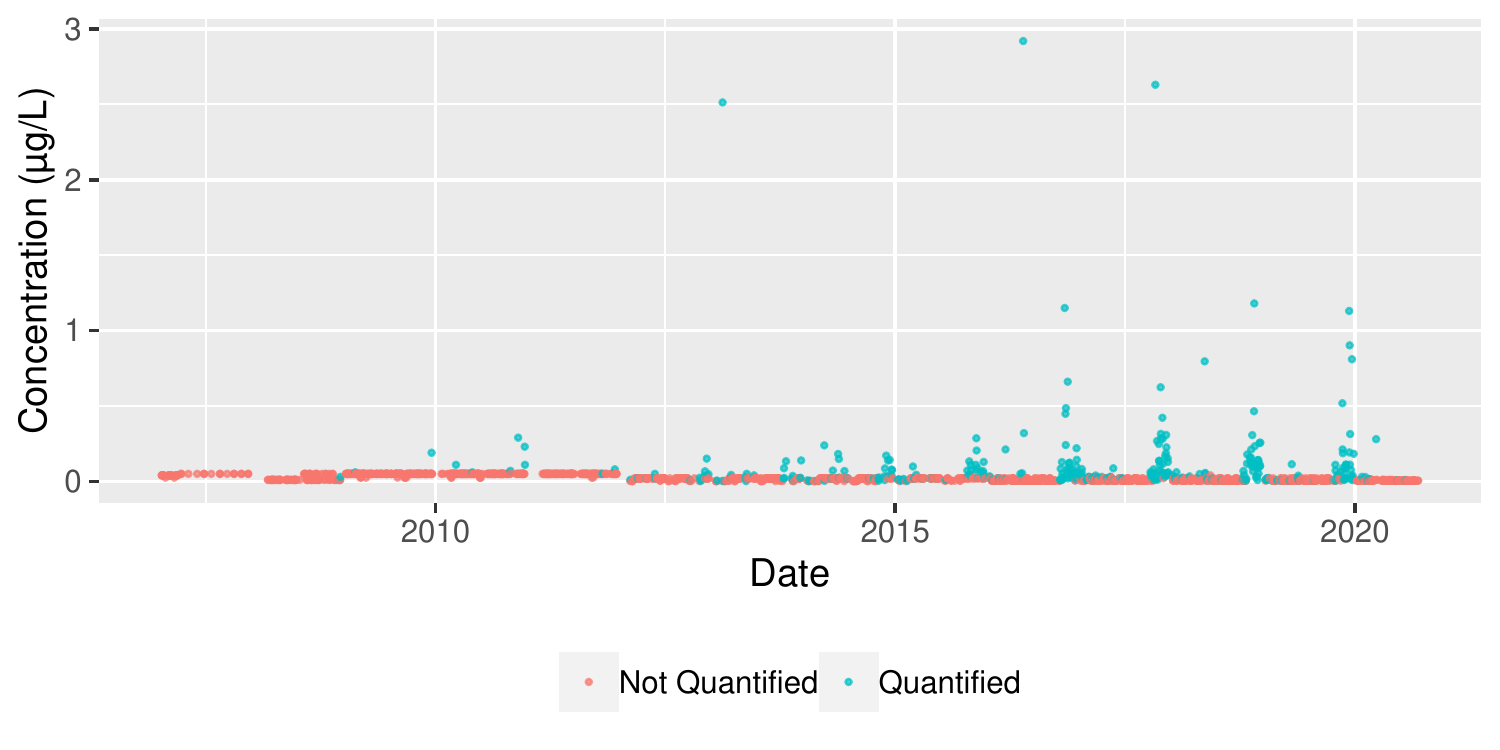}
  \caption{Plot of daily maximum concentrations}
  \label{time:serie}
\end{figure}

The coarse representation $\overline{\mathcal{D}}$ of the monitoring data $\mathcal D$ is obtained by computing the maximum daily values across the available stations. This yields the time series illustrated in Figure \ref{time:serie}. The aggregated series contains 1,808 values, among which 19.86\% are quantified. 

One may note here that despite the aggregation process, the coarse series remains irregularly sampled, and that for about two thirds of the days included in the studied time span, no measurements were made. 

\subsection{Graphical representation of the station network}

The stations network $G=(V,E)$ introduced in Section \ref{subsection:graph} is built using the hydrographic map of the Centre-Val de Loire region. Indeed, once the monitoring stations are geo-localized through their GPS coordinates, one still has to compute the edges between them, as well as the associated weights. 

For the data at hand, edges are determined using the river network. A database provided by the French National Institute of Geographic and Forest Information (IGN) \citep{IGN:BD:TOPO} contains a fine-grained description of rivers, encoded as sequences of hydrographic sections (or river sections). River sections are segments with constant geographic and hydrographic attributes. 

The procedure used for computing the edges in the stations network based on the river network may be summarised as follows:

\begin{enumerate}
    \item One starts by building a river network $R=(S,H)$, where the vertices $S$ are made of the connecting points between the river sections, and the edges $H$ contain all sections. Each edge is thus naturally weighted by the length (in meters) of the corresponding river section. 
    \item Each monitoring station $v_i$ in $V$ is assigned to the closest node $\tilde s_i$ in the river network $R$, by minimizing the geographical distance between the station $v_i$ and all connecting points
    \begin{equation*}
     \tilde s_i=\min_{s\in S} d(v_i, s).
    \end{equation*}
    \item Given two stations $v_i,v_j \in V$ and their associated connecting points $\tilde s_i,\tilde s_j \in S$, an edge will be generated between $v_i$ and $v_j$ if there exists at least one path between $\tilde s_i$ and $\tilde s_j$. Furthermore, the weight associated to an edge $(v_i,v_j)$ is equal to the length of the shortest path between $\tilde s_i$ and $\tilde s_j$.
\end{enumerate}

One may notice at this point that the above procedure may result into an unconnected graph, with several connected components. 

\begin{figure}[ht]
  \centering
  \includegraphics[]{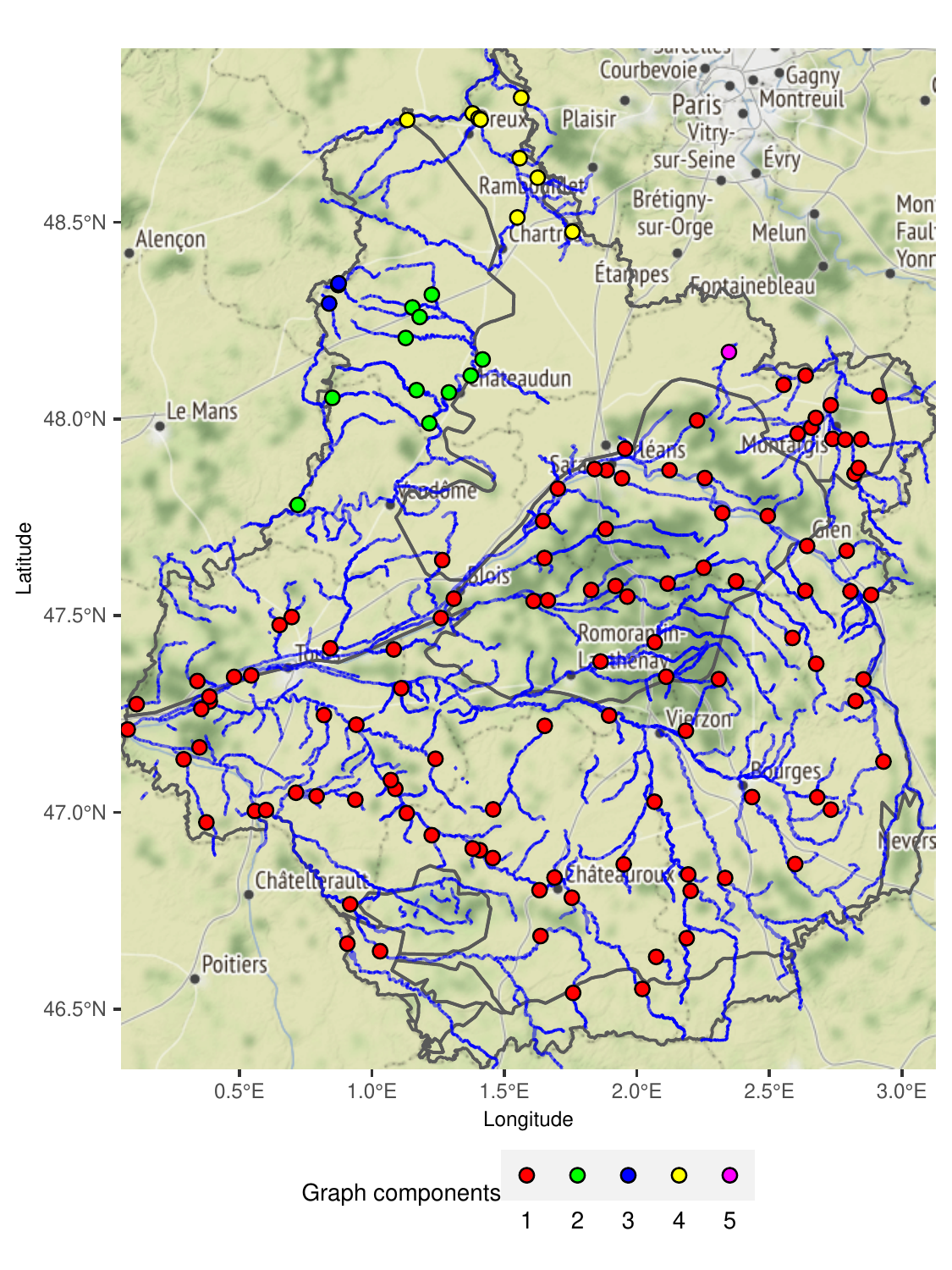}
  \caption{Connected components of 129 stations subgraph}
  \label{fig:comp}
\end{figure}

For illustration, Figure \ref{fig:comp} displays a subgraph involving 129 stations of interest that will be used later in Section \ref{section:clust129}. The resulting subgraph is not fully connected and exhibits 5 distinct connected components.  

\section{Results}\label{section:results}

\subsection{Temporal segmentation}\label{sec:time_pattern}

First, the coarse-grained time series $\overline{\mathcal{D}}$ in Figure \ref{time:serie} is segmented using the change-point detection procedure described in Section \ref{subsection:pelt}. Since the data distribution is right-skewed and shows heavy tails, a Weibull distribution is selected for the parametric distribution $Q$ (see Appendix \ref{appendix:weibull} for technical details). 

Under the assumption above, one should fit Weibull distributions within each stationary temporal segment. In order to limit the number of parameters to estimate, and also to avoid numerical issues rapidly induced by the large number of censored data, the shape parameter $\sigma$ in the Weibull distributions is supposed not to vary with the change-points. $\sigma$ is thus constant throughout the series, and is estimated globally under a stationary hypothesis. The only parameter supposed to be varying at each change-point is therefore the rate of the Weibull distribution, say $\lambda$. From the application point of view, this simplifying assumption corresponds to the hypothesis that the differences in usage and diffusion of the prosulfocarb among the different users is captured by the shape parameter, and should not vary much over time. On the contrary, the overall average usage of prosulfocarb varies, and this dependency is captured by changes in the rate parameter. 

Hence, after computing the MLE of $\sigma$, $\hat \sigma_{MLE}$, over the whole time series, change-points and rate parameters over each temporal segment are estimated by minimizing the cost function in Equation \ref{eq:cost-fct-sigma-hat}. Let us remark here that the estimated value of the shape parameter is $\hat \sigma_{MLE}=0.3$. This confirms the data has a heavier tail than an exponential distribution ($\sigma$=1), and that the assumption of using Weibull distributions for our data is appropriate.

In the change-point detection procedure, the penalty value for the PELT algorithm was calibrated using a large range of values explored according to the CROPS algorithm. The range, inspired by the BIC criterion, was set to $[\frac{\log(K)}{5},5\log(K)]$ where $K$ is the number of daily maximum concentrations available, here $K = 1,808$. Note that when using the BIC penalty in change point detection, the penalty term written in section \ref{subsection:pelt} becomes : $\beta_K(L+1)D = \frac{D}{2}\log(K)(L+1) = \frac{1}{2}\log(K)(L+1)$. The range chosen allows to screen an interval of penalties containing the BIC penalty.

\begin{figure}[ht]
  \centering
  \includegraphics[]{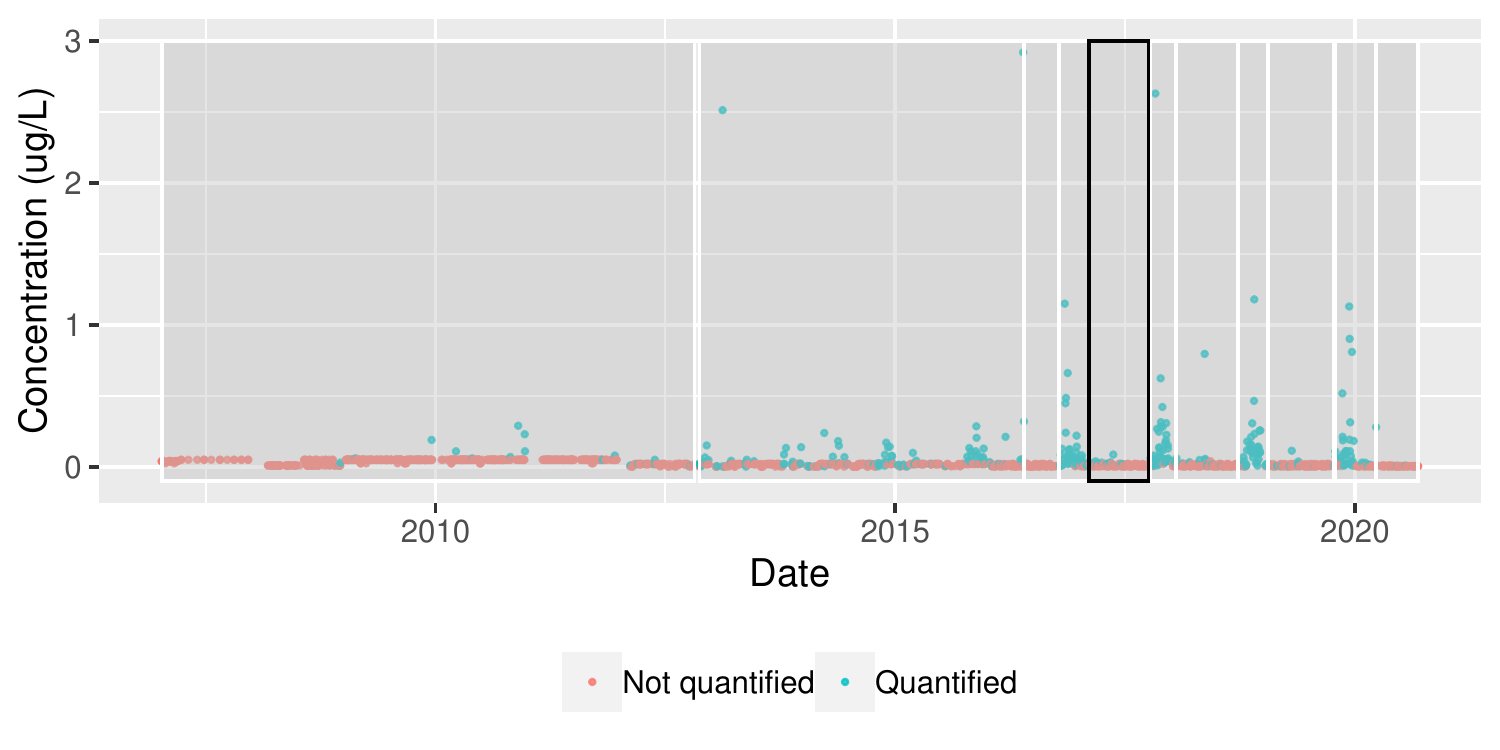}
  \caption{Best segmentation found by the change-point detection procedure with CROPS-based penalty tuning. The dates of the breaks are : October 25, 2012; May 25, 2016; October 13, 2016; February 7, 2017; September 14, 2017; January 19, 2018; October 5, 2018; January 28, 2019; October 11, 2019; March 25, 2020. The black rectangle corresponds to the selected temporal segment in section \ref{section:clust129}}
  \label{fig:seg}
\end{figure}

The penalty calibration procedure resulted in 14 different segmentations, with a number of change-points ranging from 1 to 26. The best segmentation is selected using the elbow method, as illustrated in Figure \ref{fig:elb1} in Appendix \ref{section:elb}. This amounts to a temporal segmentation with $\hat L=10$ change-points, illustrated in Figure  \ref{fig:seg}.

According to Figure \ref{fig:seg}, the usage of prosulfocarb in Centre-Val de Loire shows different patterns throughout time. Before 2016, most of the values are not quantified, and there are almost no change-points detected. Starting with 2016, two regimes of pesticide usage appear to emerge, and correspond respectively to the periods of intensive usage of prosulfocarb and to the off-peak periods. Indeed, the starting dates of the peak periods coincide with the season where the substance is spread, which is Autumn. The emergence of this two-regime pattern, alternating high concentration values during the peak periods and low concentration values during the off-peaks, is correlated with an important increase in the prosulfocarb sales as shown in Figure \ref{fig:sale} in Appendix \ref{section:sale}. 

\subsection{Spatial segmentation}\label{section:clust129}

The second step of the analysis consists in the spatial segmentation using the graph-based clustering on the monitoring stations network. Because not all monitoring stations are active during a temporal segment, the clustering procedure described in Section \ref{section:spaceclust} is slightly modified: spatial clustering is applied to the subgraph induced by the active stations only.

For illustration, let us focus on a specific temporal segment. An off-peak period, spanning between February 7, 2017 and September 14, 2017 was selected. This period was identified as a homogeneous temporal segment by the change-point detection procedure.  This period is highlighted by the black rectangle in Figure \ref{fig:seg}.

\begin{figure}[ht]
  \centering
  \includegraphics[]{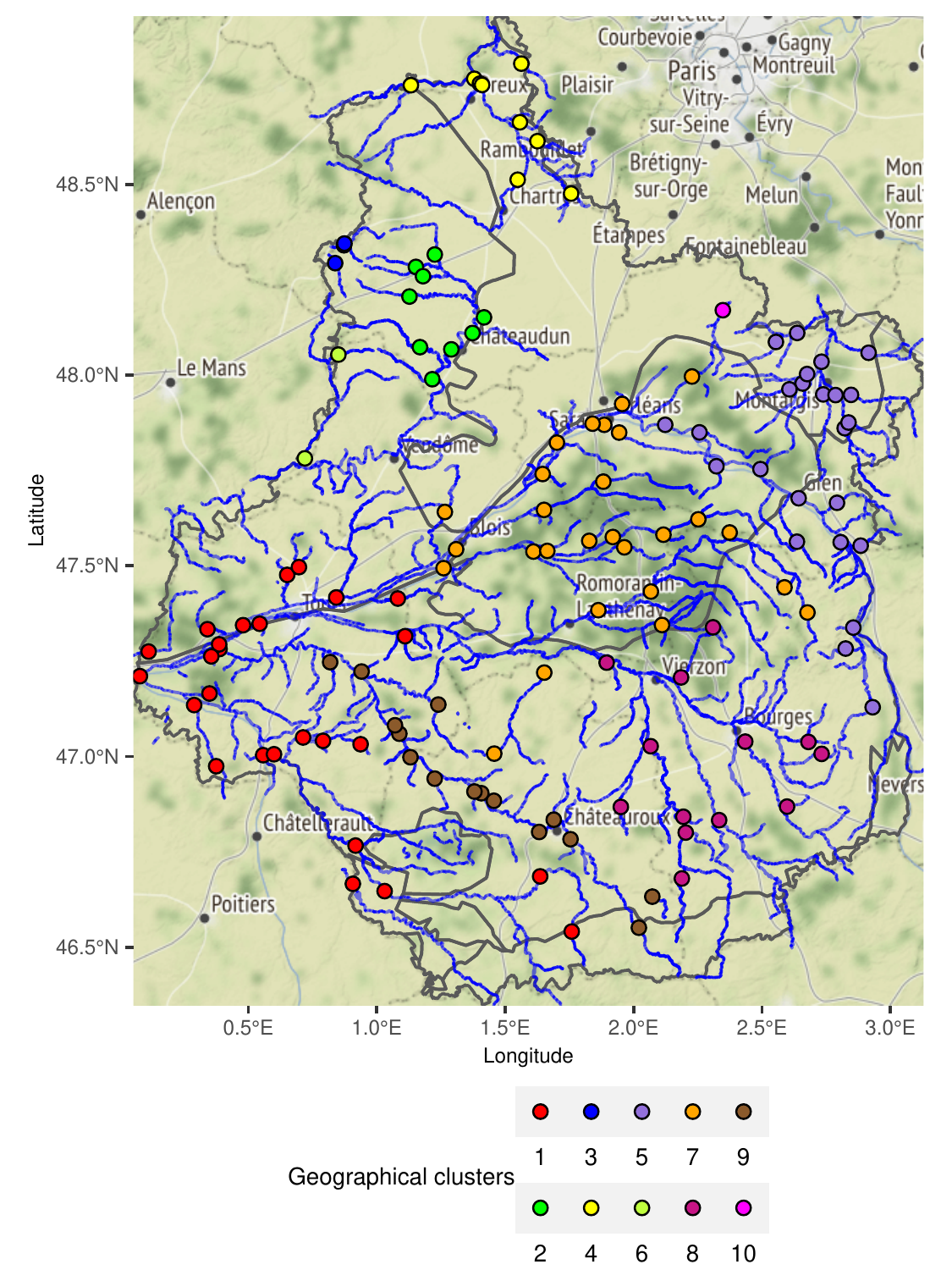}
  \caption{Optimal clustering representation. Hydro-ecoregions boundaries are given by the grey lines.}
  \label{fig:graph}
\end{figure}

During the selected period, 129 monitoring stations only produced at least one measure. The spatial clustering algorithm was applied with a number of potential clusters varying between 5 and 35 (the minimum number of clusters is equal to the number of connected components in the subgraph induced by the active stations). The optimal number of clusters was selected using the elbow method applied to the inertia curve. According to this criterion, illustrated in Figure \ref{fig:elb2} in Appendix \ref{section:elb}, the best solution is made of a 10-clusters configuration. The spatial segmentation for the selected off-peak period is illustrated in Figure \ref{fig:graph}.

To check the relevance of the homogeneity assumption formulated in section \ref{section:spaceclust}, we computed the within cluster average empirical pairwise Wasserstein distance and observe that for 8 clusters out of 10, this indicator is less than 0.0015, whereas the global average pairwise Wasserstein distance for the 129 stations is 0.003. This suggests that the distance chosen for our station graph is indeed a good proxy of the homogeneity in the concentration space. Additional comments can be made when we look at the geography of the region. Some clusters are overlapping with hydro-ecoregions. Hydro-ecoregions are geographic entities in which hydrographic ecosystems share common characteristics. The criteria defining them combine properties of geology, terrain and climate \citep{wasson:hal-02580774}. The borders of those regions are drawn in grey in Figure \ref{fig:graph}. This ensures that the substances will have homogeneous dispersion properties on these clusters (see clusters 2, 3, 4, 6 and 10 for instance). As expected the biggest component in Figure \ref{fig:comp} is the most segmented. Some among them are easy to identify, for instance clusters 9 corresponds to the Indre river. Cluster 1 is identified as the most western part of the Loire and its tributaries mainly the Vienne and the Creuse rivers. Clusters 5,7 and 8 are a little bit harder to identify. If one look closely at the map of the region, there is a high presence of small channels all across this part of the region. 

\subsection{Anomalous cluster identification}\label{subsection:anomalous}

\begin{figure}[ht]
  \centering
  \includegraphics[]{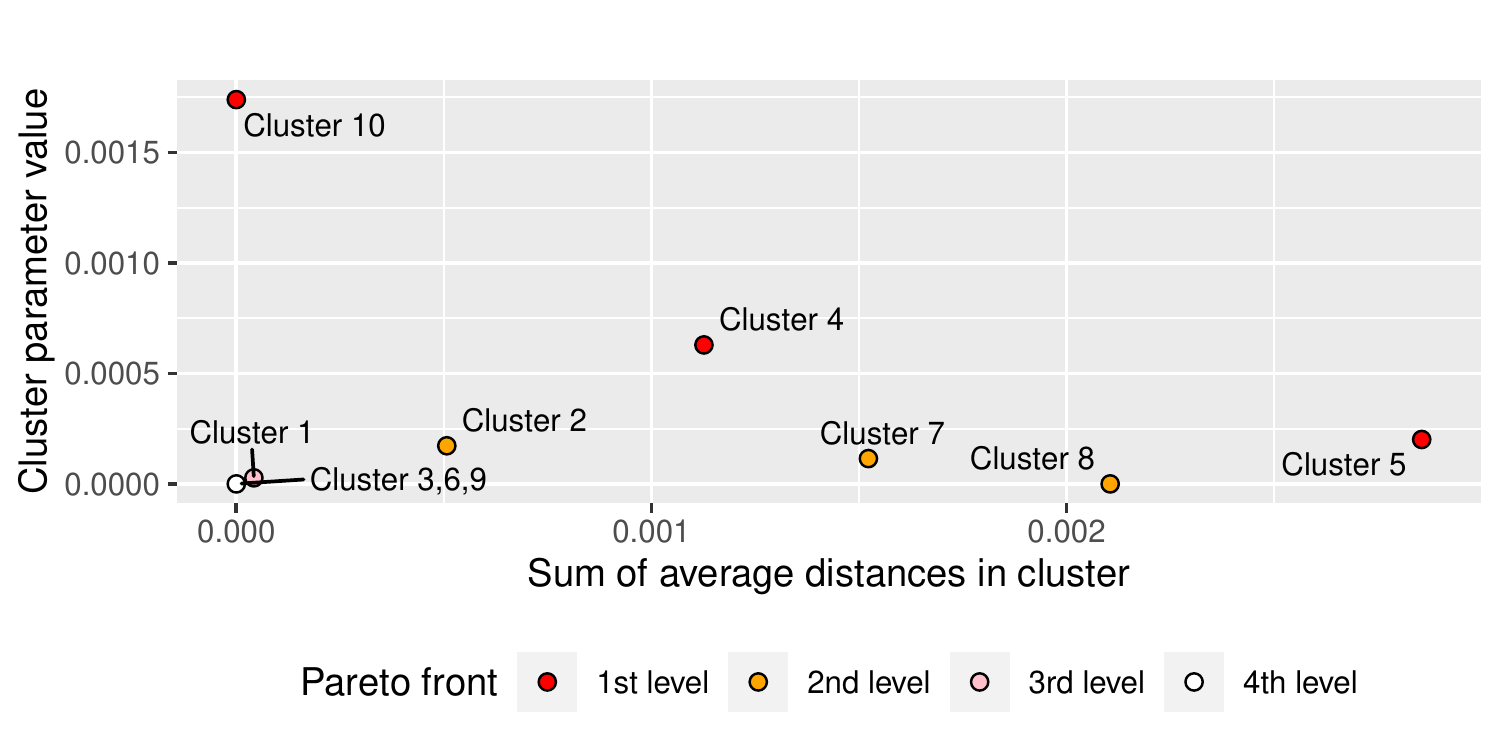}
  \caption{Pareto front displayed on the stations map}
  \label{fig:pareto}
\end{figure}

Following the methodology proposed in \ref{section:anomaly}, the scaling parameter $\lambda_k$ of the aggregated data of each spatial cluster found in Section \ref{section:clust129} was estimated. The statistics $\bar{I}_k$ was set to $1/\hat{\lambda}_k$. The Pareto front involving the two descriptors $\bar{W}_k$ and $\bar{I}_k$ was computed. It led to the cluster ranking displayed in Figure \ref{fig:pareto} using the \emph{rPref} package \citep{RJ-2016-054}. We recall that the selected time segment corresponds to a period of non-use of prosulfocarb. From this it can be deduced that finding quantified measurements of the substance during this period is an anomaly. Three clusters stood out with a Pareto front level of 1. Among them we can find from left to right on Figure \ref{fig:pareto} : 
\begin{itemize}
    \item a cluster which has recorded a very high concentration value but is composed of one station only, it corresponds to cluster 10 in Figure \ref{fig:graph}. Four measurements were made at that station among which one was quantified. It leads to a higher quantification rate than in the other clusters and high value of $1/\hat{\lambda}_k$.  
    \item a cluster where one can find different profiles of stations. Some of them recorded high concentration values while the others did not provide any quantified measurement. It corresponds to cluster 4 in Figure \ref{fig:graph}. Nine stations compose this cluster. These stations performed 68 measurements among which five were quantified. The quantified measurements can be found in three different stations. 
    \item a cluster composed of 24 stations that recorded 99 samples. This cluster recorded the maximum level of concentration during the selected time segment. Only six measurements were quantified and they are distributed over three stations. It corresponds to cluster 5 in Figure \ref{fig:graph}. The low rate of quantification leads to a low value of $1/\hat{\lambda}_k$ but the heterogeneity between concentration profiles is higher given that the station that recorded the maximum is very abnormal.
\end{itemize}

\begin{figure}[ht]
  \centering
  \includegraphics[]{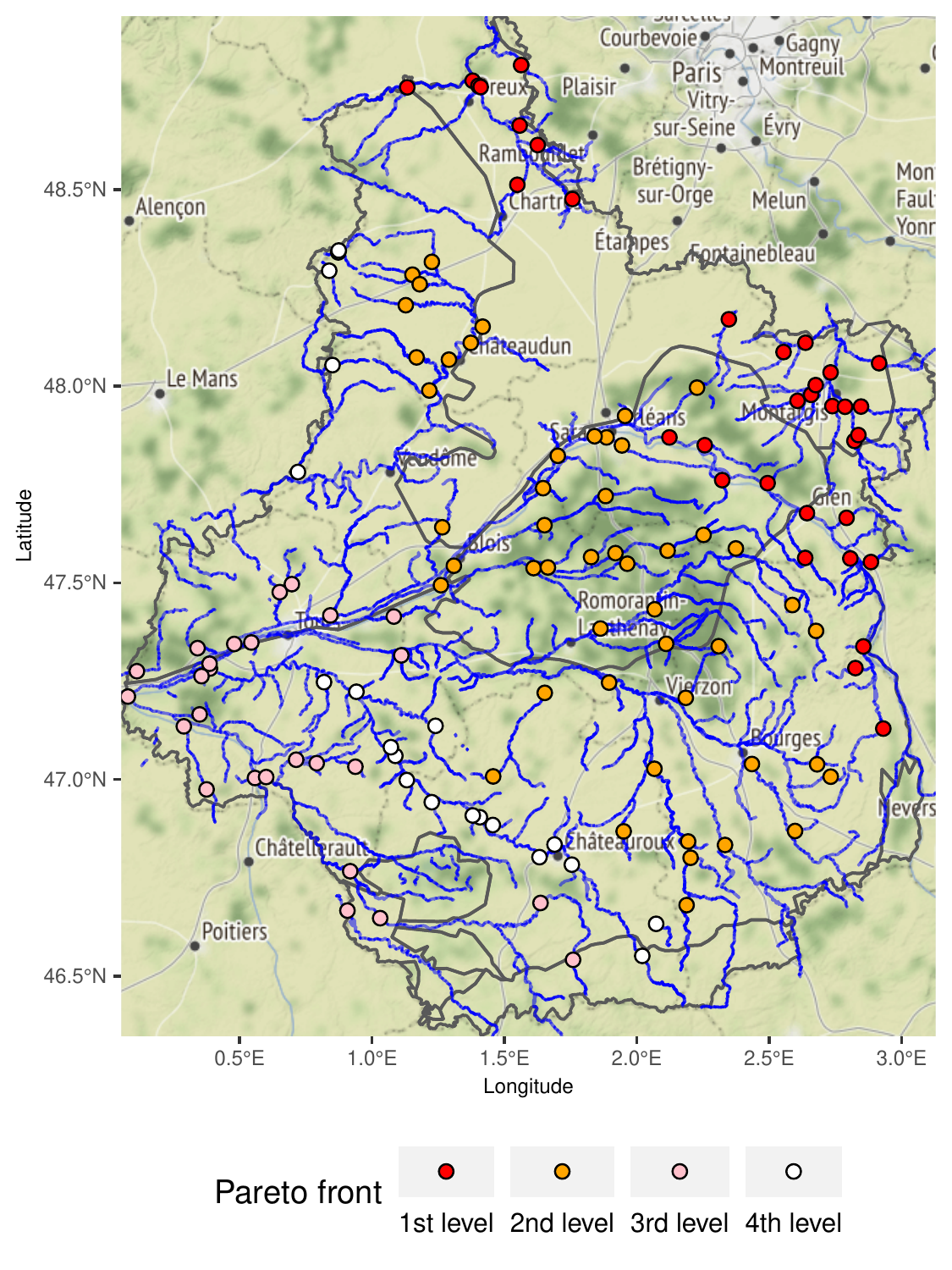}
  \caption{Pareto front displayed on the stations map}
  \label{map:pareto}
\end{figure}

Figure \ref{map:pareto} displays the Pareto front levels on the station map. 

It is interesting to note that the Pareto front level is not uniformly distributed in the region. The three anomalous clusters are located in the east of the region. It could be related to the agricultural practices and land use. For the sake of the argument, we present in Appendix \ref{section:crops} the map of barley and wheat crops in Centre-Val de Loire. In future works, we shall investigate the spatial correlation between anomalous clusters and areas with high concentration of these crops. 

\section{Conclusion}

A new methodology for investigating abnormal signals in pesticide concentration data was introduced in the present manuscript. It takes into account both the left-censored data distribution due to quantification limits, and the spatio-temporal heterogeneity due to measurements made at different stations with irregular frequency. The case study illustrated above yields promising results. Indeed, change-point detection shows clear temporal patterns and identifies periods of intensive prosulfocarb use. Focusing on specific temporal segments and combining those with the spatial clusters allows to characterise concentrations homogeneity, and peak values present in each cluster. Spatio-temporal patterns may be thus highlighted using Pareto front levels, with contiguous clusters, for instance, emerging in the Eastern part of Centre Val de Loire. However, this paper also shows that the non-regular and non-standardized collection of concentration data requires the development of a complex methodology. Therefore, comparison with existing work on spatiotemporal signal analysis is difficult \citep{Hamdi2021}. The exploratory aspect of this approach also makes comparison with existing methods difficult. This issue should encourage the various analytical laboratories performing the measurements to standardize station instrumentation and time their sampling rhythm to obtain regular time steps and homogeneous data precision. These first results are encouraging and open up new perspectives for improvements and future developments. This application may allow some operational advances in chemical analysis by a government agency. In this particular case, we indicated that Anses changed the prosulfocarb's approval in October 2017. The time segment we study in this paper precedes this period, covering the period from February to October 2017. In subsequent years, we observe a seasonal signal. For example, we could look at the other off-peak time segments in subsequent years to see if there is still as much prosulfocarb in the periods of non-use. This would help the agency measure the effectiveness of the market approval change.

In Section \ref{section:results}, we observe that the number of samples of each cluster is an important factor that affects the detection of anomalies. In fact, the detection of cluster 10 is not due to the observation of an anomalous or important quantity, but to a strong quantification of the samples. This quantification results from the fact that there are only four measured values among which one only is quantified. Since this result is not particularly satisfactory, there are several paths and ideas to tackle this problem. First, one could imagine changing the nature of the abnormality criteria. For example, one could simply look at the average of the Wasserstein distances of the immediate neighbors in the station graph. This would allow one to include more local information in the analysis of a cluster. Second, one could also change the structure of the station graph. Based on additional information, such as that found in studies of the distribution of pesticides in the air, one could edit the edges of the station graph. For example, one could connect stations that are linked by parcels of the same crop. This would have the effect of grouping components that are not connected in the current graph. The weight of these new edges could be determined by the numerous studies of pesticide dispersal in other environmental compartments (land use, air, groundwater ...etc) during spraying \citep{Payraudeau2011,Tong2002,ROZEMEIJER2007695}. Another option to perform the whole procedure is to couple change point detection with clustering and perform joint spatiotemporal segmentation instead of treating temporal and spatial heterogeneity separately. It may also be mentioned that it would be interesting to look at a multivariate approach. Co-occurrence effects between different substances are well studied \citep{Schreiner2016,Baas2016}. A new modeling on this subject could bring a better understanding of the field results.

\bibliographystyle{apacite}
\bibliography{special-issue}

\appendix

\section{Change-point detection in left-censored Weibull distributions}\label{appendix:weibull}
We instantiate in this section the general model and methodology presented in the Section \ref{subsection:pwsm} and Section \ref{subsection:pelt} in the particular case of the Weibull distribution. 

If $Y$ is a random variable distributed according to a Weibull distribution with a rate parameter $\lambda$, a shape parameter $\sigma$, the p.d.f. of $Y$ is given by 
\begin{equation*}
f_Y(y; \lambda, \sigma)=\indic{y\geq 0}\sigma\lambda\left(\lambda y\right)^{\sigma-1}e^{-\left(\lambda y\right)^\sigma},
\end{equation*}
where $\indic{}$ is the indicator function. The c.d.f. of $Y$ is given by
\begin{equation*}
F_Y(y; \lambda, \sigma)=\indic{y\geq 0}\left(1-e^{-\left(\lambda y\right)^\sigma}\right).
\end{equation*}
Then if $Y$ is subjected to left censoring with a deterministic threshold $q$, its likelihood function is given by
\begin{equation*}
L(\lambda, \sigma; y)=\left(\sigma\lambda\left(\lambda y\right)^{\sigma-1}e^{-\left(\lambda y\right)^\sigma}\right)^{\indic{y>q}}\left(1-e^{-\left(\lambda q\right)^\sigma}\right)^{\indic{y=q}}.
\end{equation*}
This likelihood function does not provide closed form estimators for $\lambda$ and $\sigma$ but any classical continuous optimisation algorithm can be used to obtain reasonable approximation of those estimators. (as in the \texttt{fitdistrplus} R package \citep{delignette2015} used in the present work). 

For change point detection, we use a fixed and global shape parameter and a segment specific rate parameter. More precisely, for a segment $[\eta_{l-1}+1, \eta_{l}]$ we use the rate parameter $\lambda_l$ (and the global shape parameter $\sigma$). Then under the assumptions of Section \ref{subsection:pwsm}, the negative log likelihood of $\maxy_{\eta_{l-1}+1}, \ldots, \maxy_{\eta_l}$ is given by
\begin{multline}
-\mathcal L_Q(\lambda_l, \sigma; \maxy_{\eta_{l-1}+1}, \ldots, \maxy_{\eta_l}) = {}-
\sum_{k=\eta_{l-1}+1}^{\eta_l}\indic{\maxy_k=\overline{q}_k}\log\left(1-e^{-\left(\lambda_l \overline{q}_k\right)^\sigma}\right)\\
+\sum_{k=\eta_{l-1}+1}^{\eta_l}\indic{\maxy_k>\overline{q}_k}\left((\lambda_l \maxy_k)^{\sigma}-\log (\sigma \lambda_l)-(\sigma-1)\log (\lambda_l \maxy_k) \right),
\end{multline}
where the $\overline{q}_k$ are the known censoring thresholds. 

We denote $\widehat{\sigma}$ the shape parameter estimated globally. Then the cost function is given by  
\begin{equation}
\mathcal{C}_{\widehat{\sigma}}(L,  \bm{\eta}, \bm{\lambda};   \overline{\mathcal{D}})=\sum_{l=1}^{L+1}  -\ln \mathcal L_Q(\lambda_l, \widehat{\sigma}; \maxy_{\eta_{l-1}+1}, \ldots, \maxy_{\eta_l}) + \beta_K(L+1),
\label{eq:cost-fct-sigma-hat}
\end{equation}
where we use $\widehat{\sigma}$ as a subscript to emphasize the fact that the shape parameter is not estimated in each segment. 

This cost function can be analyzed as proposed in Section \ref{subsection:pelt} by plugging in the MLE estimator of $\bm{\lambda}$ obtained by maximizing $\mathcal L_Q(\lambda_l, \widehat{\sigma}; \maxy_{\eta_{l-1}+1}, \ldots, \maxy_{\eta_l})$  for $l=1,\ldots,L+1$. Notice that despite the fact $\widehat{\sigma}$ is fixed during those calculations, we still need to rely on an optimisation algorithm as no closed form formula can be derived to compute the rate parameters under censoring. 

\section{Assessing the homogeneity of the measurements of the stations}\label{appendix:wasserstein}
\begin{figure}
    \centering
    \includegraphics{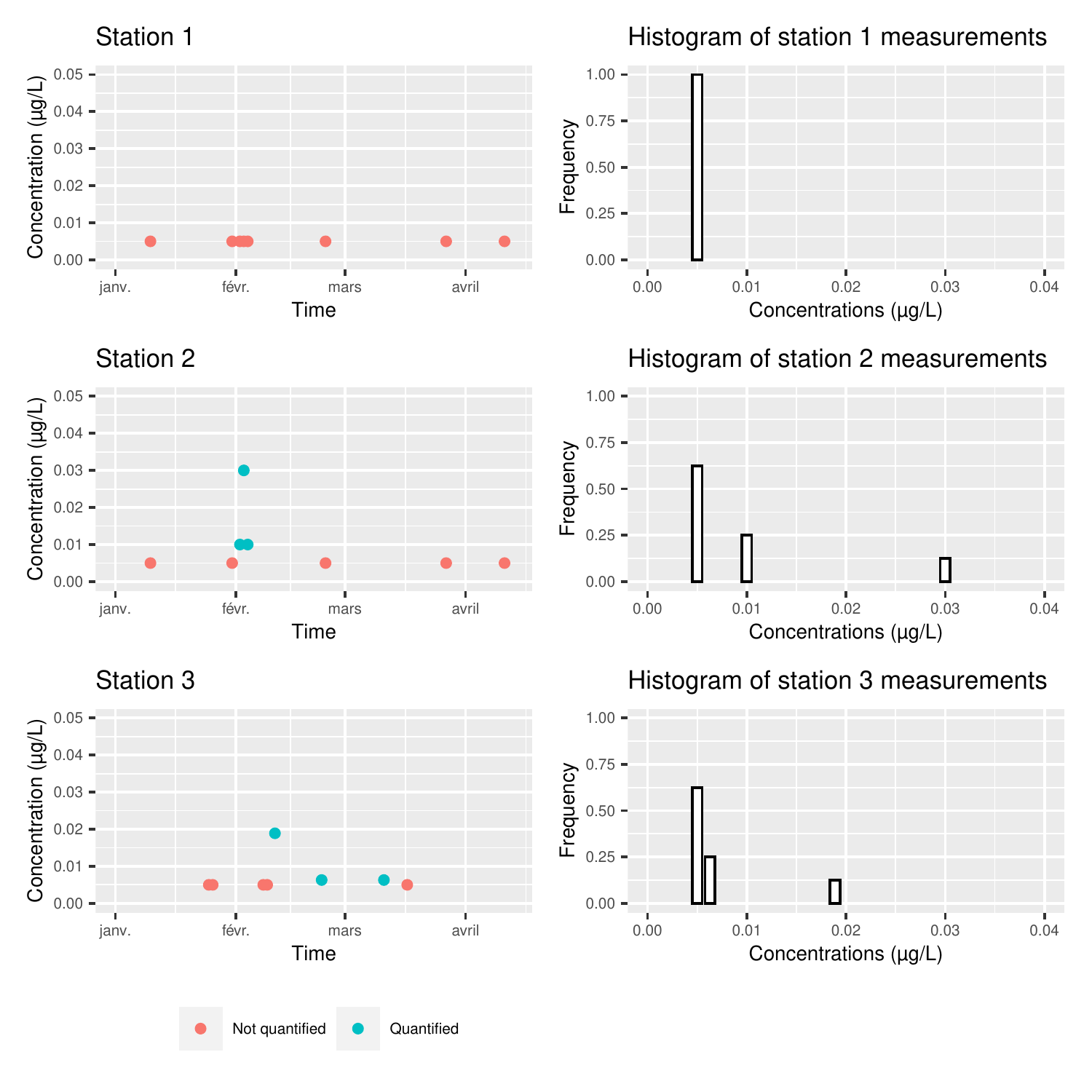}
    \caption{Example of three stations data. The data were simulated.}
    \label{fig:ex_dist}
\end{figure}

The Wasserstein distance was chosen over the Kolmogorov-Smirnov or the Jensen-Shannon metric. It has the advantage of integrating in the distance calculation both the differences between the probabilities of observing different values but also the distances between those values. This is a critical point which is illustrated on a simple simulated example provided by Figure \ref{fig:ex_dist}. We show here three monitoring stations that have quite different behaviors. Those different behaviors are obvious both on in the temporal representation and in the histograms. However, the Kolmogorov-Smirnov distance between stations 1 and 3 is equal to the Kolmogorov-Smirnov distance between stations 1 and 2. This distance cannot capture the fact that station 2 recorded higher concentration values than station 3. On the contrary, the Wasserstein distance between stations 1 and 3 is smaller than the Wasserstein distance between stations 1 and 2. 

Computing information theoretic distances/dissimilarities such as the Jensen-Shannon divergence requires estimating densities for the distributions observed at the stations. As noted earlier, few concentration records (and even fewer quantified ones) are available at the level of a station and within a time period. Therefore, density estimations based on such a small number of observations are unreliable.

\clearpage

\section{Supplementary figures}

\subsection{Regional map of crops}\label{section:crops}
\begin{figure}[ht]
    \centering
    \includegraphics{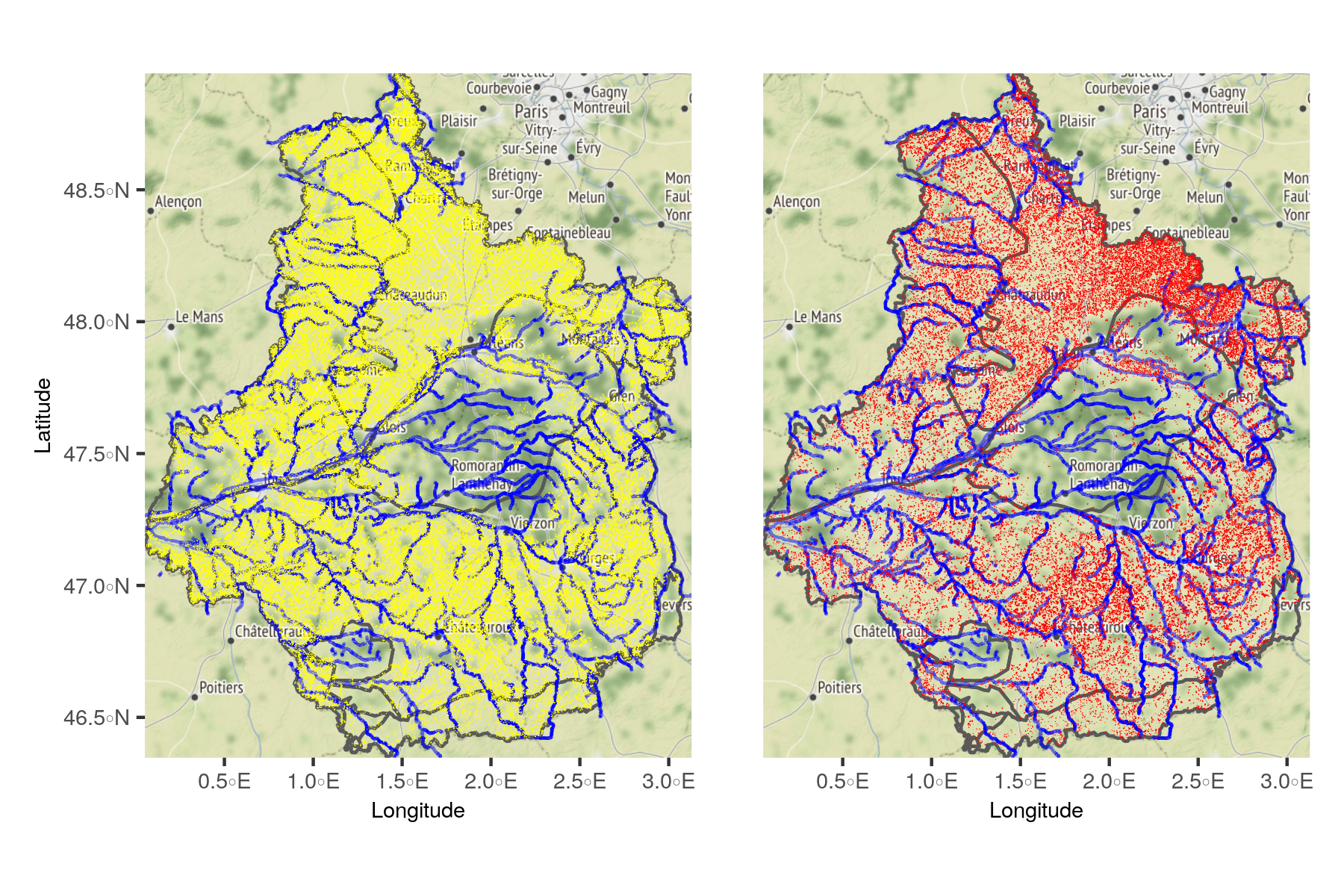}
    \caption{Wheat (in yellow) and barley (in red) crops location in Centre-Val de Loire}
    \label{fig:crops}
\end{figure}

The regional map of crops provided in Figure \ref{fig:crops} have been produced using data from the \emph{registre parcellaire graphique} produced by the IGN \citep{IGN:RPG}.

\subsection{Prosulfocarb sales}\label{section:sale}
\begin{figure}[ht]
  \centering
  \includegraphics[]{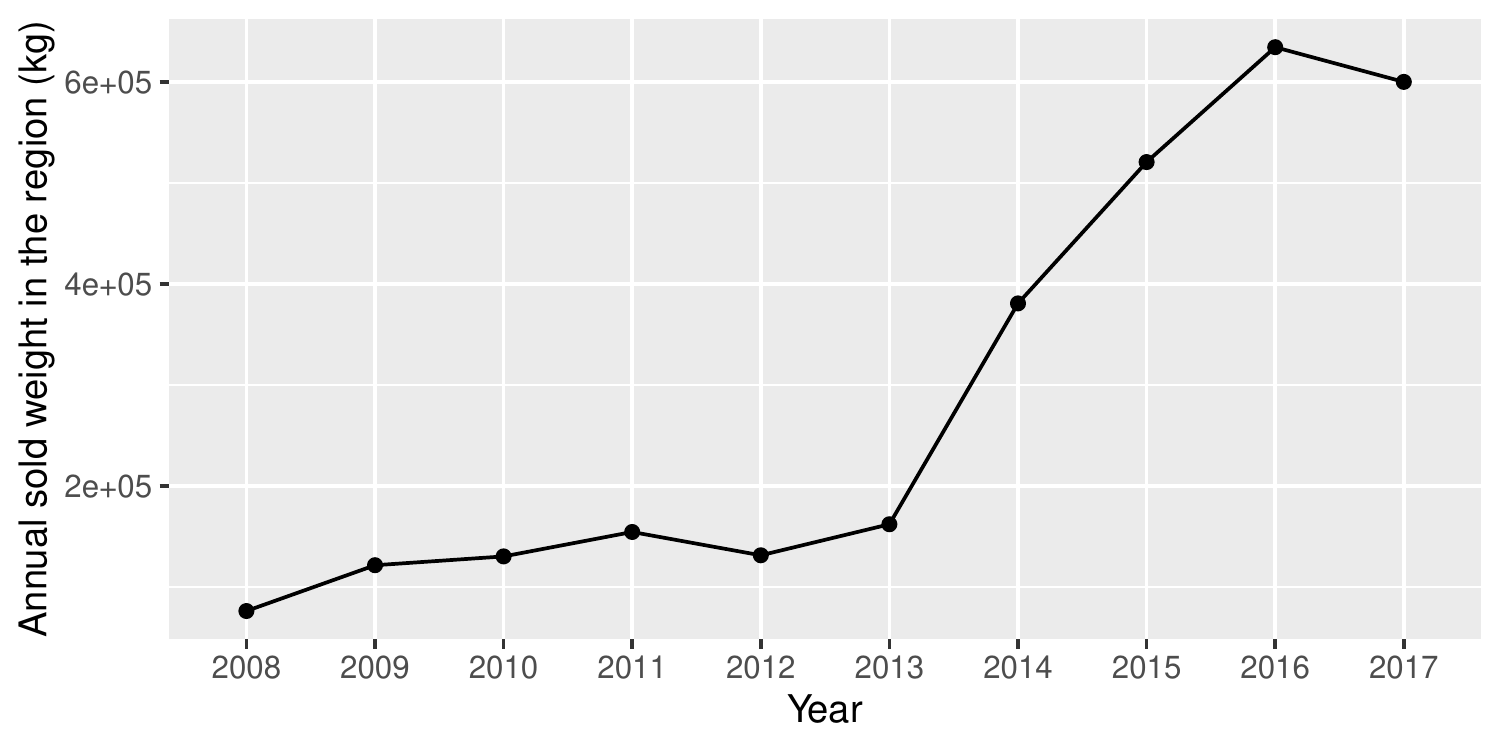}
  \caption{Prosulfocarb sales between 2008 and 2017 in the Centre-Val de Loire region}
  \label{fig:sale}
\end{figure}

Prosulfocarb sales figures used to build Figure \ref{fig:sale} are made available by the \emph{Système d'information sur l'eau} \citep{BNVD}.

\subsection{Elbow methods}\label{section:elb}

\begin{figure}[ht]
  \centering
  \includegraphics[]{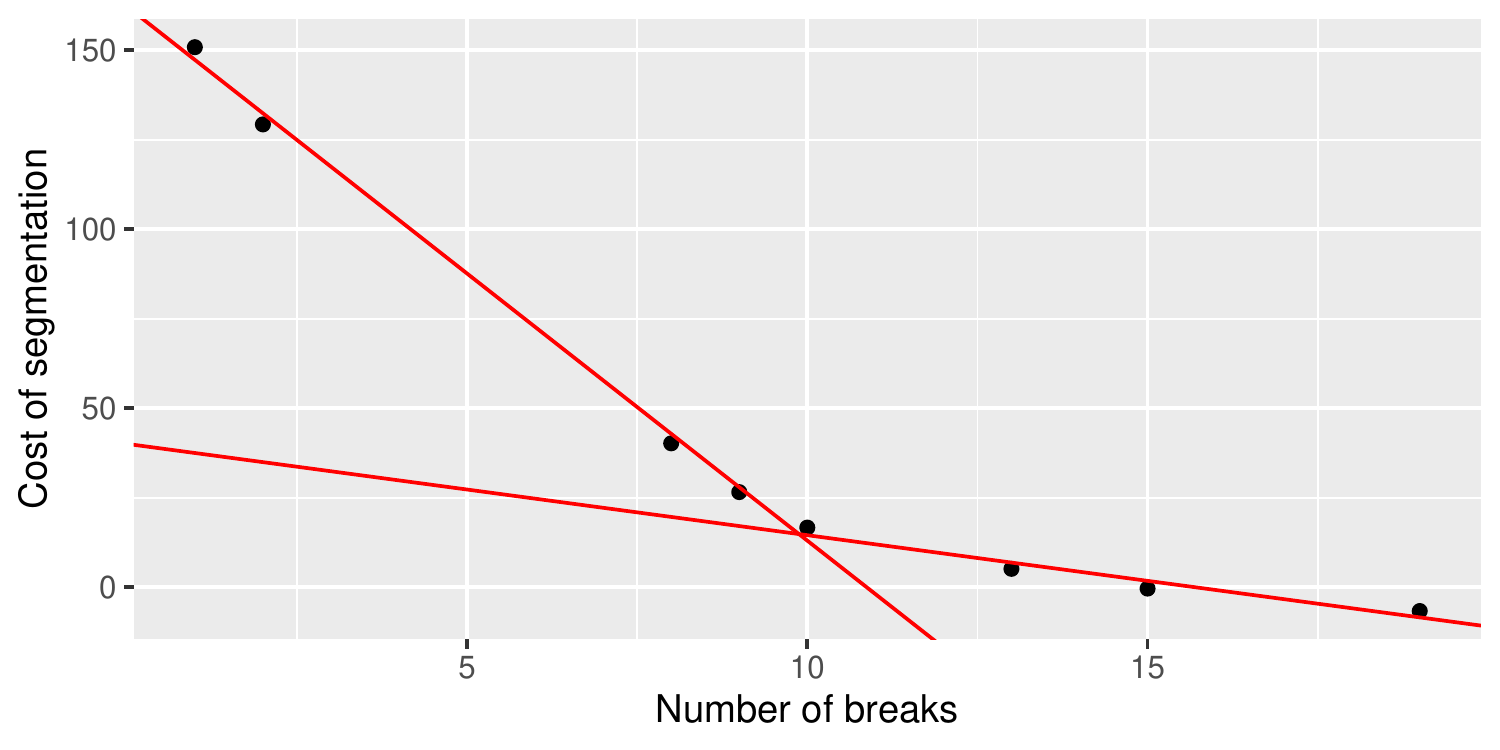}
  \caption{Elbow method illustration for the temporal segmentation. The optimal penalty value according to this approach is 8.36.}
  \label{fig:elb1}
\end{figure}

\begin{figure}[ht]
  \centering
  \includegraphics[]{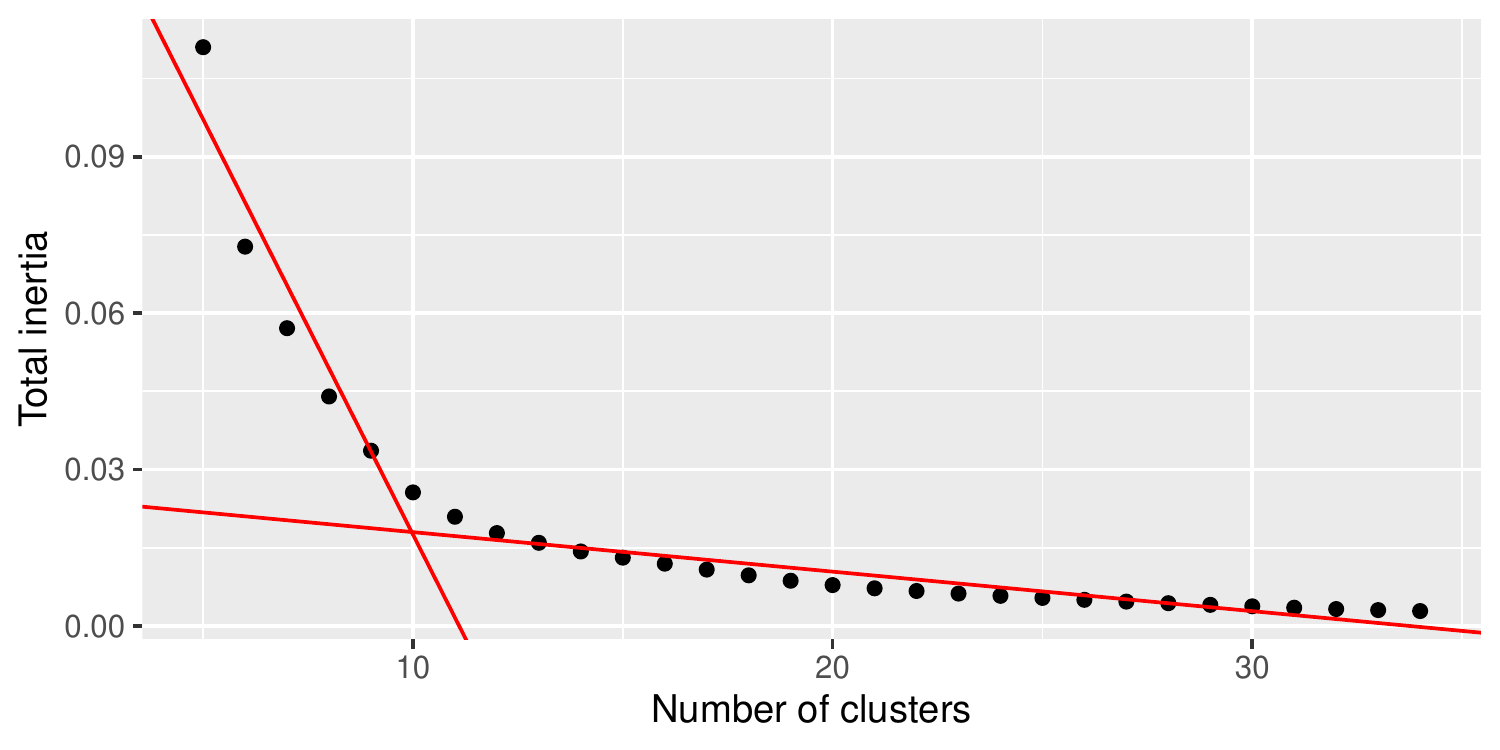}
  \caption{Elbow method illustration for the spatial clustering. The optimal number of clusters according to this approach is 10.}
  \label{fig:elb2}
\end{figure}

\end{document}